\documentclass[12pt]{JHEP3}
\usepackage{mathrsfs}
\usepackage{amsmath,amssymb}
\usepackage{epsfig}
\input epsf

\def\x'{\mathaccent 19 x}
\def\y'{\mathaccent 19 y}
\def\n'{\mathaccent 19 n}
\def\u'{\mathaccent 19 u}

\def\et'{\mathaccent 19 \eta}
\def\th'{\mathaccent 19 \theta}
\def\lam'{\mathaccent 19 \lambda}
\def\varet'{\mathaccent 19 \vartheta}
\def\rh'{\mathaccent 19 \rho}
\def\ph'{\mathaccent 19 \phi}
\def\xb'{\mathaccent 19 {\bar{x}}}

\def\l{{\lambda}}

\def\be{\begin{equation}}
\def\ee{\end{equation}}

\newcommand{\bea}{\begin{eqnarray}}
\newcommand{\eea}{\end{eqnarray}}

\def\a {\alpha}
\def\b {\beta}
\def\s {\sigma}
\def\pa {\partial}

\def\g {\gamma}
\def\p{\phi}
\def\la{\label}

\def\ov{\over}

\def\D{\Delta}

\newcommand{\alg}[1]{\mathfrak{#1}}
\newcommand{\su}{\alg{su}}
\newcommand{\sls}{\alg{sl}}

\newcommand{\AdS}{{\rm  AdS}_5\times {\rm S}^5}

\def\L{\mathscr L}

\newcommand{\sfrac}[2]{{\textstyle\frac{#1}{#2}}}

\preprint{ {\tt hep-th/0510208}\\ {\tt ITP-UU-05/47} \\ {\tt
SPIN-05/32}\\
{\tt AEI-2005-160}}

\title{Uniform Light-Cone Gauge for Strings in ${\bf \rm AdS}_5
\times {\bf \rm S}^5$: Solving $\su(1|1)$ Sector}

\author{
Gleb Arutyunov$^{a}$\footnote{e-mail: G.Arutyunov@phys.uu.nl,
frolovs@aei.mpg.de} \footnote{Also at Steklov Mathematical
Institute, Moscow } and Sergey Frolov$^{b}$\footnote{ Also at
SUNYIT, Utica, USA and Steklov Mathematical Institute, Moscow.}
\\
\\
$^{a}$ {\it Institute for Theoretical Physics and Spinoza Institute,
Utrecht University \\
~~3508 TD Utrecht, The Netherlands}\\
$^{b}$ {\it Max-Planck-Institut f\"ur Gravitationsphysik,
Albert-Einstein-Institut}\\
~~Am M\"uhlenberg 1, D-14476 Potsdam, Germany\\
}

\abstract{We introduce a uniform light-cone gauge for strings
propagating in $\AdS$ space-time. We use the gauge to analyze
strings from the $\su(1|1)$ sector, and show that the reduced
model is described by a quadratic action for two complex fermions.
Thus, the uniform light-cone gauge allows us to solve the model
exactly. We analyze the near BMN spectrum of states from the
$\su(1|1)$ sector and show that it correctly reproduces the $1/J$
corrections. We also compute the spectrum in the strong coupling
limit, and derive the famous $\lambda^{1/4}$ asymptotics. We then
show that the same string spectrum can be also derived by solving
Bethe ansatz type equations, and discuss their relation to the
quantum string Bethe ansatz for the $\su(1|1)$ sector. }

\begin{document}

\newpage

\renewcommand{\thefootnote}{\arabic{footnote}}
\setcounter{footnote}{0}
\section{Introduction}
In recent years we have received an impressive evidence that
integrability might provide a key concept to get more insight into
the complicated dynamics of the large $N$ gauge and string
theories, and, in particular, into the AdS/CFT duality conjecture
\cite{M}. Indeed, the ${\cal N}=4$ SYM in the large $N$ limit is
conjectured to be an integrable model
\cite{Minahan:2002ve,Beisert:2003tq}, at least in a certain
asymptotic approximation. Integrability allows one to formulate
the corresponding Bethe ansatz whose solutions encode the spectrum
of the model. A nice and distinguished feature of this approach is
that at many instances the Bethe equations can be solved exactly
or used indirectly to make a comparison with the dual string
theory. There has been a lot of discussion in the recent
literature concerning construction and applications of the Bethe
ansatz to the ${\cal N}=4$ SYM theory, we refer the reader to the
comprehensive reviews \cite{Beisert:2004ry}.

\medskip

The sigma-model describing Type IIB superstrings propagating in
the $\AdS$ space-time \cite{MT} is also classically integrable
\cite{Bena:2003wd}. However, due to the large number of dynamical
variables and their involved interactions the quantization problem
looks highly non-trivial. An interesting insight into the quantum
theory can be gained by studying an expansion around the so-called
plane-wave limit \cite{Metsaev:2001bj} where the string theory
simplifies dramatically but still allows for a non-trivial
comparison to the dual gauge theory \cite{Berenstein:2002jq}.

\medskip

Perhaps one of the main outcomes of string integrability is that
the string sigma-model admits a rich variety of explicit soliton
solutions \cite{FT,AFRT}. In fact the whole classical (finite-gap)
spectrum is encoded into a set of certain integral (Bethe type)
equations \cite{Kazakov:2004qf} supported on the corresponding
algebraic curves \cite{Beisert:2005bm}. Finally, the (quantum)
gauge and (classical) string theories reveal a certain interesting
similarity between their integrable structures that can be
manifested either through the study of infinite towers of
conserved charges \cite{Arutyunov:2003rg,Das:2004hy,Alday:2005gi}
or, equivalently, by comparing the corresponding  Bethe equations
\cite{Kazakov:2004qf,Beisert:2004hm}. A complementary approach to
quantum string  based on the knowledge of a quantum integrable
sigma-model in the infinite volume was suggested in
\cite{Mann:2005ab}.

\medskip

Recently, the knowledge of the classical string Bethe equations
\cite{Kazakov:2004qf} together with the asymptotic gauge theory
Bethe ansatz \cite{Beisert:2004hm} allowed us to conjecture a
novel Bethe ansatz \cite{Arutyunov:2004vx} which is supposed to
capture the leading {\it quantum dynamics} of strings in $\AdS$.
We will refer to the corresponding construction as the quantum
string Bethe ansatz.  Conjectured originally for the so-called
$\su(2)$ sector,  it has been generalized to other sectors
\cite{Staudacher:2004tk} and, finally, to the whole superstring
sigma-model \cite{Beisert:2005fw}. Classical spinning strings, the
$1/J$ corrections to energies of the plane-wave states, the famous
$\lambda^{1/4}$ strong coupling asymptotics, all these limiting
cases can be derived from the quantum string Bethe ansatz. As was
shown very recently \cite{SZZ,BT}  the  ansatz seems to be capable
to incorporate the $1/J$ corrections to classical spinning
strings. Quite intriguing, it also admits interpretation in terms
of integrable long-range spin chains \cite{Beisert:2004jw} and
naturally emerges in the study of the plane-wave matrix models
\cite{Fischbacher:2004iu}. However, in spite of all these
remarkable developments it remains unclear how the quantum string
Bethe ansatz could arise upon  quantization of strings beyond the
semi-classical approximation\footnote{Possible sources of the
corrections to the gauge/string Bethe ans\"ate have been recently
discussed in \cite{Ambjorn:2005wa}.}.

\medskip

On the other hand, it is known that string theory admits
consistent truncations to smaller sectors which contain in
particular string states dual to operators from the corresponding
closed sectors of gauge theory. In  gauge theory a closed sector
is an invariant subspace of composite operators on which the
action of the dilatation operator closes. Studying the mixing
problem within a closed sector provides certain simplifications,
e.g., in formulating the corresponding Bethe ansatz, etc. One can
try to apply a similar idea to string theory. Instead of dealing
with the complicated dynamics of the whole model one can {\it
consistently truncate} the classical string equations to a smaller
set of fields and further study their dynamical properties. One
can also try to construct the quantum theory of a truncated sector
although it is not a priori guaranteed that this theory will have
a certain relation to the actual quantum string: the procedures of
truncation and quantization are not expected to commute. Thus it
is of interest to look at this problem: It might help to
understand the origin and the range of validity of the quantum
string Bethe ansatz as well as interrelation between truncation
and quantization procedures.

\medskip

From all varieties of the closed sectors \cite{Beisert:2004ry} on
the gauge theory side
 the so-called $\su(1|1)$ sector seems
particularly attractive. In the ${\cal N}=1$ language it contains
composite operators made of two Yang-Mills elementary fields, $Z$
and $\Psi$, where $Z$ is the complex scalar from a scalar
supermultiplet and $\Psi$ is the Weyl fermion from the gaugino
supermultiplet. The $\su(1|1)$ symmetry group transforms $Z$ and
$\Psi$ into each other. In this sector the dilatation operator
\cite{Beisert:2004ry} and the corresponding asymptotic Bethe
ansatz \cite{Staudacher:2004tk} are known up to three-loop order
of perturbation theory; at one loop the dilatation operator just
coincides with the Hamiltonian of the free lattice fermion
\cite{Callan:2004dt}. The coherent state description of the
$\su(1|1)$ sector with its further comparison to string theory was
considered in \cite{ST}.

\medskip

In our previous work \cite{AAF} we have found the consistent
truncation of the classical superstring theory to the  $\su(1|1)$
sector. We have further removed all unphysical degrees of freedom
by fixing the so-called uniform gauge \cite{KT,AF}: The
world-sheet time $\tau$ was identified with the global AdS time
$t$, while the momentum of an angle variable $\phi$ of ${\rm S}^5$
was declared to be equal to the Noether charge $J$ corresponding
to translations of $\phi$. The space-time energy $E$ of the string
coincides in this approach with the world-sheet Hamiltonian $H$
which is a function of the charge $J$:
$$
E=H(J)\, .
$$
The physical degrees of freedom are two complex fermions which can
be organized in a single world-sheet Dirac fermion. The resulting
theory appears to be a new {\it non-trivial  interacting theory of
the 2-dim massive Dirac fermion}. It is integrable because the
consistent reduction can be carried over for the Lax
representation of the original sigma-model. Finally, we used the
corresponding Hamiltonian to derive the $1/J$ correction to the
energies of the plane-wave states and found a perfect agreement
with the results by \cite{Callan:2004ev,McLoughlin:2004dh}.

\medskip

To proceed with quantization one has to first identify  the action
and angle variables for the classical model. This is rather
non-trivial in our present setting because the Lagrangian of the
reduced theory is apparently complicated and involves terms up to
six order in fermions and their derivatives.

\medskip

In this paper we will solve and find the semi-classical spectrum $E(J)$
of our interacting theory in a way which bypasses direct
diagonalization of the interacting Hamiltonian. The basic idea is
to fix reparametrization invariance by choosing a gauge  most
suitable for computing the spectrum of $E$. Quite remarkably,
there exists a gauge choice which linearizes equations of motion!
In the following we will refer to this gauge as the {\it uniform
light-cone gauge}. In the light-cone coordinates
$x_{\pm}=\sfrac{1}{2}(\phi\pm t)$ this gauge consists in fixing
$x_+=\tau$ and $p_+=P_+={\rm const}$, where $p_+$ is the momentum
conjugate to $x_-$. The world-sheet Hamiltonian $H$ and the
parameter $P_+$ are now related to the global charges of the model
$E$ and $J$ as
$$
H=E-J \, , ~~~~~~~~P_+=E+J\, .
$$
Since $H$ itself is a certain function of $P_+$ we get an equation
$$
E=J+H(E+J)\, ,
$$
which can be solved for the energy $E\equiv E(J)$. It appears that
the Hamiltonian corresponding to this gauge choice is just
the quadratic Hamiltonian for two complex fermions! Thus,
in the uniform light-cone gauge the theory becomes free and the
spectrum of $H$ is trivially computed.

\medskip

It is worth emphasizing that in this way we solved our reduced
model exactly, i.e., without assuming anything about the range of
$J$. Taking $J$ to be large we can easily construct the $1/J$
expansion of the energy. The leading order in this expansion is
the energy of a plane-wave state, while at the subleading order we
again reproduce the $1/J$ correction found in
\cite{Callan:2004ev,McLoughlin:2004dh}. Remarkably, to obtain this
correction in our approach we need only free fermions.
The subleading $1/J^2$ correction turns out to be an
analytic function of $\lambda'=\lambda/J^2$, while the recent
studies \cite{BT,SZ} of the $1/J$ correction to energies of
classical spinning strings suggest the appearance at this order of
non-analytic terms. This possible mismatch can be attributed
to the fact that the $\su(1|1)$ sector is not closed in quantum
string theory.

\medskip

It is also easy to analyze the strong coupling expansion,
$\lambda\to\infty$, of our exact result. Again we reproduce the
leading asymptotic $\lambda^{1/4}$, and find a disagreement of the subleading
$\lambda^{-1/4}$ correction with the quantum string Bethe ansatz predictions.

\medskip

The paper is organized as follows. In section 2 we explain the
uniform gauge approach which is then applied in section 3 to the
$\su(1|1)$ sector. We show that the gauge-fixed Hamiltonian is
that for free massive fermions. In section 4 we compute the
spectrum of the model and analyze the near-plane wave and strong
coupling expansions. We also discuss generalization to the case of
non-vanishing winding. In section 5 we establish a relation
between the energy spectrum we found and the quantum string Bethe
ansatz. In Conclusion we discuss the consequences of our results
and open problems. Finally, in the appendix A we compute the
$1/J^2$ and $1/\sqrt{\lambda}$ corrections by using the quantum
string Bethe ansatz and in appendix B we discuss the Lax pair
which arises upon fixing the uniform light-cone gauge.

\section{Uniform Light-Cone Gauge}
In this section we introduce the uniform light-cone gauge for
strings propagating on a target manifold. This gauge generalizes
the standard phase-space light-cone gauge of \cite{Goddard:1973qh}
to a curved background \cite{Metsaev:2000yu}. It belongs to
the class of gauges used to study the dynamics of strings in
$\AdS$ \cite{KT,AF}.
\medskip

We denote the time coordinate of the manifold by $t$, and assume
that the manifold possesses a U(1) isometry realized by shifts of
an angle variable $\phi$. To impose the uniform light-cone gauge
we also assume that the string sigma-model action is invariant
under shifts of the time coordinate $t$ and the angle variable
$\phi$, with all the other bosonic and fermionic fields being
invariant under the shifts. This means that the string action does
not have an explicit dependence on $t$ and $\p$ and depends only
on the derivatives of the fields. An example of such a string
action is provided by the Green-Schwarz superstring in $\AdS$
where the metric can be written in the form \bea \nonumber ds^2 =
f_a(z) dt^2 + f_s(y) d\p^2 + g^a_{ij}(z)dz^idz^j +
g^s_{ij}(y)dy^idy^j \ . \eea Here $t$ is the global time
coordinate of ${\rm AdS}_5$, $\p$ is an angle of ${\rm S}^5$, and
$z^i$ and $y^i$ are the remaining coordinates of  ${\rm AdS}_5$
and ${\rm S}^5$, respectively. Strictly speaking, the original
Green-Schwarz action presented in \cite{MT} contains fermions
which are charged under the U(1) transformations generated by the
shifts of $t$ and $\p$. However, it is possible to redefine the
fermions and make them neutral in the same way as it was done in
\cite{AAF} for fermions from the $\su(1|1)$ sector.

\medskip

The invariance of the string action under the shifts leads to the
existence of two conserved currents, $E^\a$ and $J^\alpha$, and
two conserved charges
$$
E = \int_0^{2\pi} {{\rm d}\s\ov 2\pi}\, E^0\ ;\qquad J=
\int_0^{2\pi} {{\rm d}\s\ov 2\pi}\, J^0\ .
$$
It is clear that the charge $E$ is the target space-time energy,
and $J$ is the total U(1) charge of the string. It is well-known
that the time components, $E^0$ and $J^0$, of the abelian charges
are just equal to the momenta conjugate to the coordinates $t$ and
$\p$:\footnote{We assume that the kinetic term in the Hamiltonian
form of the string action has the form $ - p_t\dot{t} + p_\p
\dot{\p} +\cdots$, with the negative sign in front of $p_t$. }
$$
p_t \equiv E^0\;\qquad \qquad p_\p\equiv J^0\ .
$$
To impose the uniform light-cone gauge we introduce the light-cone
coordinates: \bea \la{lcc} &&t = x_+ - x_-\ ,\quad \p = x_+ + x_-\
, \quad p_t = {1\ov 2}(p_+ + p_-) \ ,\quad p_\p = {1\ov 2}(p_+ -
p_-)\\\nonumber &&x_+ ={1\ov 2}(\p +t)\ ,\quad x_- ={1\ov 2}(\p
-t)\ , \quad p_+ = p_\p+p_t\ ,\quad p_- = p_t -p_\p \ . \eea In
terms of the light-cone coordinates the kinetic term takes the
form \bea \la{kin} - p_t\dot{t} + p_\p \dot{\p} = -p_-\dot{x}_++
p_+ \dot{x}_-\ . \eea Then we fix the uniform light-cone gauge by
the conditions \bea \la{ulc} x_+ = \tau + {m\ov 2}\s\ ,\quad p_+ =
P_+ = E + J\ {\rm ~is\ a\ constant}\ . \eea

 The integer number $m$ is the winding number
which appears because the coordinate $\p$ is an angle variable
with the range $0\le\p\le 2\pi$. It is clear from (\ref{kin}) that
in this gauge the 2-dim Hamiltonian is identified with the
integral over $\s$ of the momenta $p_-$: \bea \la{H2d} H =
\int_0^{2\pi} {{\rm d}\s\ov 2\pi}\, p_- = E-J\ . \eea In the
AdS/CFT correspondence the space-time energy $E$ of a string state
is identified with the conformal dimension $\D$ of the dual CFT
operator: $E\equiv \D$. Since the Hamiltonian $H$ is a function of
$P_+=E+J$, the relation (\ref{H2d}) gives us a nontrivial equation
on the energy $E$. Computing the spectrum of $H$ and solving the
equation (\ref{H2d}) would allow us to find conformal dimensions
of dual CFT operators.

\section{The $\su(1|1)$ sector}
In this section we use the uniform light-cone gauge to analyze the
string theory on $\AdS$ reduced to the $\su(1|1)$ sector. It was
shown in \cite{AAF} that the sector contains the scalars $t$ and
$\phi$, and two complex fermions $\vartheta_3$ and $\vartheta_8$.
The Lagrangian\footnote{We are using the notations from
\cite{AAF}. The reader can consult \cite{AAF} for details of the
reduction.} of the reduced model can be written in the form \bea
\label{Lagrangian} \L
&=&\sfrac{\sqrt{\lambda}}{2}\gamma^{\tau\tau}\left(\dot{t}^2-
\dot{\phi}^2+\sfrac{i}{2}(\dot{t}+\dot{\phi}) \zeta_{\tau}
-\sfrac{1}{2}(\dot{t}+\dot{\phi})^2 \Lambda
\right) \\
\nonumber
&+&\sfrac{\sqrt{\lambda}}{2}\gamma^{\sigma\sigma}\left({t'}^2-
{\phi'}^2+\sfrac{i}{2}({t'}+{\phi'}) \zeta_{\sigma}
 -\sfrac{1}{2}(t'+\phi')^2 \Lambda
\right)\\
\nonumber
&+&\sqrt{\lambda}\gamma^{\tau\sigma}\left(\dot{t}t'-\dot{\phi}\phi'
+\sfrac{i}{4}(\dot{t}+\dot{\phi}) \zeta_{\sigma}
+\sfrac{i}{4}(t'+\phi') \zeta_{\tau}
-\sfrac{1}{2}(\dot{t}+\dot{\phi})(t'+\phi')\Lambda \right) +
\L_{\rm wz}\, .\eea Here the Wess-Zumino term has a remarkably
simple form $(\kappa = \pm\sqrt{\l}/2)$ \bea \label{WZ} \L_{\rm
wz}&=\frac{\kappa}{2}\Omega_{\tau}(t'+\phi')-\frac{\kappa}{2}
\Omega_{\sigma}(\dot{t}+\dot{\phi})
 \, ,
 \eea
and for various fermionic contributions we use the concise
notations \bea \label{constr1}
\begin{array}{lll}
 \zeta_{\tau}=\vartheta_i\dot{\vartheta}^i+\vartheta^i\dot{\vartheta}_i \, ,
 ~~&~~
\Omega_{\tau}=\vartheta_3\dot{\vartheta}_8+\vartheta_8\dot{\vartheta}_3
-\vartheta^3\dot{\vartheta}^8-\vartheta^8\dot{\vartheta}^3\, ,
~~&~~
\Lambda=\vartheta_i\vartheta^i\, ,\\
\zeta_{\sigma}=
\vartheta_i{\vartheta'}^i+\vartheta^i{\vartheta'}_i \, , ~~&~~
\Omega_{\sigma}=\vartheta_3\vartheta'_8+\vartheta_8\vartheta'_3
-\vartheta^3\vartheta'^8-\vartheta^8\vartheta'^3 \, . ~~&~~
\end{array}
\eea
It is important to mention that the periodicity condition for the fermions
$\vartheta_3$ and $\vartheta_8$ depends on the winding number $m$.
If $m$ is even the
fermions are periodic, and if $m$ is odd they are anti-periodic. The dependence
appears because one makes the original periodic fermions neutral under the
shifts of $t$ and $\p$ by means of a field redefinition, and this
induces the change in
the periodic condition, see \cite{AAF} for details.\footnote{An equivalent change
of boundary conditions was also found in the analysis of the
spectrum of fluctuations around a
multi-spin circular string \cite{FT3}. }

\medskip

In \cite{AAF} the action (\ref{Lagrangian}) was studied by
imposing the phase-space uniform gauge $t =\tau$, $p_\p=J$, where
$p_{\phi}$ is the canonical momentum conjugate to the angle
variable $\phi$. It was shown that the gauge-fixed action arising
in this way defines an integrable model of a massive {\it
interacting} Dirac fermion described by the following Lagrangian
\bea \label{lorentz}
\L&=&J\Big[-1-\frac{1}{2}\left(i\bar{\psi}\rho^{\a}\pa_{\a}
\psi-i\pa_{\a}\bar{\psi}\rho^{\a}\psi\right)
+\bar{\psi}\psi
\\
\nonumber
 &&~~~-\frac{1}{4}\epsilon^{\a\beta}(
\bar{\psi}\pa_{\a}\psi~\bar{\psi}\rho^5\pa_{\beta}\psi-
\pa_{\alpha}\bar{\psi}\psi~\pa_{\beta}\bar{\psi}\rho^5\psi
)
  +\frac{1}{8}\epsilon^{\alpha\beta}(\bar{\psi}\psi)^2
\pa_{\a}\bar{\psi}\rho^5
\pa_{\beta}\psi \Big]\, ,
\eea
where $\psi$ is the fermion,
and $\rho^\a$ are 2-dim Dirac matrices.
Here we will reanalyze (\ref{psL}) by
imposing the uniform light-cone gauge, and show that in this gauge
the gauge-fixed action is
a free action for the fermions $\vartheta_3$ and $\vartheta_8$.
This allows us to
solve the integrable system described
by the Lagrangian (\ref{lorentz}) in the semi-classical approximation.
We do not know however if the change of gauge leads to quantum-equivalent
models because the reduction of the string theory to the $\su(1|1)$ sector
breaks the conformal invariance that is necessary for quantum equivalence of
different gauges. It would be interesting to quantize (\ref{lorentz}) directly,
and compare its spectrum with the spectrum we find from the light-cone gauge
free fermion action.

\medskip

Introducing the light-cone coordinates (\ref{lcc})
 we rewrite (\ref{Lagrangian}) in the Hamiltonian form \bea
\nonumber \L=&-&p_-\dot{x}_+ +p_{+}\dot{x}_- -\frac{i}{4}p_+
\zeta_{\tau}+\kappa x'_+\Omega_{\tau}\\
\nonumber &-&\frac{1}{\gamma^{\tau\tau}\sqrt{\lambda}}
\Big[\frac{1}{2}p_+p_-+\frac{1}{4}p_+^2\Lambda-\frac{\kappa}{2}p_+\Omega_{\sigma}
+\frac{i}{2}\lambda x'_+\zeta_{\sigma}-2\lambda x'_+x'_-
-\lambda {x'_+}^2\Lambda\Big]\\
\label{psL} &+&\frac{\gamma^{\tau\sigma}}{\gamma^{\tau\tau}}
\Big[-\frac{i}{4}p_+ \zeta_{\sigma}+p_+ x'_--p_-x'_+ +\kappa
x'_+\Omega_{\sigma}\Big] \, . \eea

As is usual in string theory with two-dimensional
reparametrization invariance, the components of the world-sheet
metric $\g^{\a\b}$ enter the phase-space Lagrangian in the form of
the Lagrangian multipliers. Imposing the uniform light-cone gauge
(\ref{ulc}) and solving equations of motion for the components
$\gamma^{\tau\tau}$ and $\gamma^{\tau\sigma}$ we find
\bea
\la{pm}
p_- &=& \kappa\Omega_{\sigma} -{1\ov 2}P_+\Lambda\, ,\\
\la{xmp} x_-' &=& {i\ov 4}(\zeta_{\sigma} + i m\Lambda)\ .
\eea
Integrating (\ref{xmp}) over $\s$, we get the level-matching
condition
\bea
{\cal V}=\la{lm} \int_0^{2\pi} {{\rm d}\s\ov 2\pi}\, {i\ov
2}(\zeta_{\sigma} + i m\Lambda) = m\ .
\eea
As usual the
condition should be imposed on the physical states of the model.
In fact the field $x_-$ is unphysical and varying the Lagrangian
w.r.t. $p_+$ we find that it evolves according to a first-order
equation
\bea
\dot{x}_-=\frac{i}{4}(\zeta_{\tau}+2i\Lambda)\, .
\eea
The components of the world-sheet metric can be found from
the equations of motion for $p_-$ and $x_-$ and they are given by
\bea
\gamma^{\tau\tau}=\frac{1}{2}\Big(\frac{\sqrt{\lambda}
}{P_+}m^2-\frac{P_+}{\sqrt{\lambda}} \Big)\, , ~~~~~~~~~
\gamma^{\tau\sigma}&=&-\frac{m\sqrt{\lambda}}{P_+} \, .
\eea
Since
the unitarity of the model requires $\gamma^{\tau\tau} < 0$ we get
that $P_+=E+J > \sqrt{\lambda}|m|$. The origin of this condition is
easy to understand. If $m\neq 0$ the string winds around a circle,
and has the length equal to $2\pi |m|$. The energy of such a string
must be greater than the product of the string tension and its
length, and this leads to the condition. The condition shows that
the energy of a long winding string always scales as
$\sqrt{\lambda}$ \cite{AF} as opposite to  the usual $\l^{1/4}$
scaling of a short string with $m=0$.

\medskip

Finally, substituting the solutions of the Virasoro constraints to
the Lagrangian (\ref{psL}), we get the gauge-fixed Lagrangian for
strings in the $\su(1|1)$ sector \bea \la{gxL} \L = -{i\ov 4}
P_+\zeta_{\tau} + {1\ov 2}\kappa m\Omega_{\tau} - \kappa
\Omega_{\sigma}+{1\ov 2}P_+\Lambda \, .\eea Recalling the
definitions (\ref{constr1}), we see that the action is a {\it
free} action for two complex fermions! If the winding number $m=0$
then rescaling $\s$ it can be cast to the form of the action for a
free massive Dirac fermion \cite{AAF}.

\section{Spectrum}
In this section we discuss the spectrum of the model. We start
with the simplest case of the vanishing winding number $m=0$. In
this case the Lagrangian coincides with the quadratic part of the
Lagrangian obtained in \cite{AAF}, and, therefore, we can just use
the results from \cite{AAF}. It was shown there that after a
proper change of the fermionic variables (see section 7 of
\cite{AAF}) the action (\ref{gxL}) takes the form \bea
\label{lagr} {\cal L} = \sum_{n=-\infty}^\infty\,\left[ -i \left(
a^+_n\dot{a}^-_n +b^+_n\dot{b}^-_n\right) - \omega_n\left( a^+_n
a^-_n +b^+_n b^-_n\right)\right] \, , \eea where
$\omega_n=\sqrt{1+\tilde{\l} n^2}$, and we define $\tilde{\l}$ by
the formula
$$
\tilde{\l} = {4\l\ov P_+^2} = {4\l\ov (E+J)^2}\ .
$$
 In terms of the oscillators $a^\pm, b^\pm$ the level
matching condition has the usual form
\bea
\label{levmat} {\cal V}
= {2\ov P_+}\sum_{n=-\infty}^\infty\,\left(n\, a^+_n a^-_n - n\, b^+_n
b^-_n\right) = 0\ ,
\eea
and therefore the sum of $a$-modes should be
equal to the sum of $b$-modes. As was discussed in \cite{AAF}, the
SYM operators from the $\su(1|1)$ subsector are dual to states
obtained by acting by operators $a^+_n$ on the vacuum. A general
$M$-impurity state with $M=M_a+M_b$ obtained by acting by $M_a$
operators $a^+_n$ and $M_b$ operators $b^+_n$ is
\bea
\label{state} |M_a,M_b\rangle =b_{j_1}^+\ldots b_{j_{M_b}}^+\,
a_{i_1}^+\ldots a_{i_{M_a}}^+ |0\rangle \,.
\eea
It is obvious
that the 2-dim energy of this state is equal to
\bea
\la{spectrum}
H|M_a,M_b\rangle =\left(
\sum_{i=1}^M\omega_{n_i}\right)|M_a,M_b\rangle\ .
\eea
As was
discussed in section 2, the 2-dim energy of a string state is
related to the space-time energy by the formula $H = E-J$. Taking
into account (\ref{spectrum}), we get the following equation for
the space-time spectrum of string states
\bea
\la{space-time} E-J
= \sum_{i=1}^M\sqrt{1+{4\lambda n_i^2\ov (E+J)^2}}\ .
\eea
Since
all fermions are neutral under the U(1) subgroup shifting the
bosonic field $\phi$, the state (\ref{state}) carries the same $J$
units of the corresponding charge for any number of excitations
$M$. That means that an $M$-impurity string state should be dual
to the field theory operator of the form \bea \la{oper} {\rm
tr}\big( \Psi_+^{M_a} \Psi_-^{M_b}Z^{J-\frac{M}{2}} \big)+\ldots
\,, \eea where $\Psi_\pm$ are the two fermions from the gaugino
multiplet of ${\cal N}=4$ SYM, carrying the Lorentz charge
$\sfrac{1}{2}$ and $-\sfrac{1}{2}$ under one of the $\su(2)$'s
from the Lorentz algebra $\su(2,2)$. According to the AdS/CFT
correspondence, the space-time energy $E$ of a string state is
equal to the conformal dimension of the dual CFT operator, and,
therefore, solutions of
 eq.(\ref{space-time}) give us dimensions of the operators (\ref{oper}).

\medskip

In what follows we restrict our attention to the states dual to
the closed $\su(1|1)$ subsector of gauge theory. For such states
$M_b=0$, and the sum of the modes vanishes.

\subsection{Near-plane wave correction to the energy }
It is very simple to use eq.(\ref{space-time}) to compute the
$1/J$ correction to the energy of the plane-wave states from the
$\su(1|1)$ sector. All one needs to do is to introduce the
effective coupling constant $\l' = \l/ J^2$, and solve
(\ref{space-time}) in powers of $1/J$ keeping $\l'$ and $M$
finite. After a simple algebra we find \bea \la{ppw} E - J =
\sum_{i=1}^M\omega_i\left(1 - {\l'\ov 2J}
\sum_{j=1}^M{n_j^2\ov\omega_j}\right) + {\cal O}(1/J^2)\ , \eea
where $\omega_i = \sqrt{1+\lambda' n_i^2}$. Taking into account
the level-matching condition this formula can be rewritten in the
form \bea E=J&+&\sum_{i=1}^M\omega_i
-\frac{\lambda'}{4J}\sum_{i\neq j}^M\frac{n_i^2+n_j^2+2n_i^2n_j^2
\lambda'-2n_in_j\omega_i\omega_j}{\omega_i\omega_j}\, . \eea This
{\it precisely} reproduces
 the $1/J$ correction to the $M$-impurity
plane-wave states obtained in \cite{McLoughlin:2004dh, AAF} by
using nontrivial interacting Hamiltonian for fermions. Here we
reproduced the spectrum by using {\it free} fermions!

\medskip

It is clear that eq.(\ref{space-time}) can be used to compute the
$1/J^2$ and higher corrections. For the $1/J^2$ correction we find
\bea \la{ppw2} E_2 ={\l'\ov 8} \sum_{i,j,k=1}^M\omega_i
\omega_j{n_k^2(3+2\l' n_k^2)\ov \omega_k^3} + {\l'^2\ov 4}
\sum_{i,j,k=1}^M\omega_k{n_i^2n_j^2 \ov \omega_i\omega_j} \ , \eea
where $E - J = \sum_{i=1}^M\omega_i + {E_1\ov J} + {E_2\ov J^2} +
\cdots$. However, since the $\su(1|1)$ sector is not closed in
quantum string theory one should also take into account
contributions from the fields which were set to zero in the
reduction to the sector. In particular, we do not see the term
$\l'^{5/2}/J^2$ recently predicted from the analysis of the $1/J$
correction to spinning strings \cite{BT}. Moreover, as follows
from the analysis in next section, eq.(\ref{ppw2}) does not
reproduce correctly even all terms analytic in $\l$. The $1/J^2$
corrections in the $\su(2)$ sector were recently studied in
\cite{MTT} by using a properly adjusted fast-string action. Our
results suggest that to derive the action one would have to take
into account the contribution of fields that are not from the
$\su(2)$ sector.

\subsection{ Strong coupling limit }

In our derivation of eq.(\ref{space-time}) we never assumed that
$J$, and therefore $E$ must be of order $\sqrt{\l}$. That means
that we can also consider the strong coupling limit when
$\l\to\infty$ and $E\sim \lambda^{1/4}$. We need to consider the
two cases: {\it i)} $J\sim 1$, {\it ii)} $J\sim E \sim
\lambda^{1/4}$. To simplify the notations it is convenient to
introduce the radius of $S^5$: $R = \sqrt{\l}$. Then, in the first
case, $J\sim 1$, we find \bea \la{strong1} E = 2\sqrt{n R}\left(1
+ {1\ov 8R}\left({J^2\ov n}+\sum_i{1\ov |n_i|}\right) + {1\ov
8\sqrt{n}R^{3/2}}\sum_i{1\ov |n_i|} + \cdots\right)\ , \eea where
$n\equiv {1\ov 2}\sum_i |n_i|$ is the level of the string state.
To understand the physical meaning of the formula it is useful to
find $E^2$: \bea \la{strong2} E^2 = 4n R+J^2 + n\sum_i{1\ov |n_i|}
+ {\sqrt{n}\ov \sqrt{R}}\sum_i{1\ov |n_i|} + \cdots\ . \eea It is
clear from the formula that the first two terms give the usual
dispersion relation for a string state of level $n$ moving in the
$\phi$-direction with the momenta $J$. The remaining terms are the
leading $1/\sqrt{R}$ corrections. Let us also note that the
expansion goes in powers of $1/\sqrt{R} = 1/\l^{1/4}$.
\medskip

In the second case, $J\sim E \sim \lambda^{1/4}$, the
eq.(\ref{space-time}) gives \bea \la{strong3} E = \sqrt{4 n R +
J^2}\left(1 + {1\ov 2R}\left(n+{1\ov 2}j(j+\sqrt{4 n +
j^2})\right)\sum_i{1\ov |n_i|}  + {\cal O}(1/R^2)\right) ,~~~~~~
\eea where $j = J/\sqrt{R}$ is kept finite in the expansion. In
this case the expansion goes in powers of $1/R = 1/\sqrt{\l}$.

We do not expect that our formulas reproduce correctly the leading
strong coupling corrections because as we will discuss in the next
section and in the appendix A they do not match the predictions of
the quantum string Bethe ansatz.

\subsection{Non-vanishing winding number}
Here we discuss the spectrum of the model for the case of
non-vanishing winding number $m\neq 0$.
In this case one can show that there is a change of the fermions
such that the action (\ref{gxL}) takes the form
\bea
\label{lagrm}
{\cal L} = \sum_{n=-\infty}^\infty\,\left[ -i \left(
a^+_n\dot{a}^-_n +b^+_n\dot{b}^-_n\right) - \omega_n^+\, a^+_n
a^-_n - \omega_n^-\, b^+_n b^-_n\right] \, ,
\eea
where
\bea
\la{oma}
\omega_n^\pm={P_+\sqrt{P_+^2-\l m^2 + 4\l n^2} \pm2\l mn\ov P_+^2-\l m^2}\ .
\eea
Since the fermions are periodic if the winding number $m$ is even and
anti-periodic if $m$ is odd, the mode numbers $n$ in (\ref{lagrm}) are integer
or half-integer, respectively.

The space-time energy of a generic string state (\ref{state})
can be again found from the equation (\ref{H2d}) that takes the form
\bea
\la{ener3}
E - J = \sum_{i=1}^{M_a} \omega_{n_i}^+ + \sum_{i=1}^{M_b} \omega_{k_i}^-\ .
\eea
The string states must satisfy the level matching condition (\ref{lm}).
In terms of the
oscillators $a^\pm, b^\pm$ it has a rather unusual form
\bea
\label{levmat2} {\cal V} =
\sum_{n=-\infty}^\infty\,\left(c_n^+\, a^+_n a^-_n - c_n^-\, b^+_n
b^-_n\right)= m\ ,
\eea
where
\bea
\la{cnpm}
c_n^\pm = {2n P_+\pm m \sqrt{P_+^2-\l m^2 + 4\l n^2} \ov P_+^2-\l m^2}\ .
\eea
Acting by the level-matching condition on a string state (\ref{state}),
we get the following condition on the mode numbers
\bea
\label{levmat3}
\sum_{i=1}^{M_a} c_{n_i}^+ - \sum_{i=1}^{M_b} c_{k_i}^- = m\ .
\eea
It is not difficult to show by using (\ref{ener3}) and (\ref{cnpm}) that
the condition just says that the sum of $a$-modes minus
the sum of $b$-modes is equal to $mJ$:
\bea
\label{levmat4}
\sum_{i=1}^{M_a} n_i - \sum_{i=1}^{M_b} k_i = mJ\ .
\eea
For states from the $\su(1|1)$ sector we have $M_b=0$. The simplest state
is created by acting by the operator $a^+_{mJ}$ on the vacuum. In this case
eq.({\ref{ener3}) can be solved exactly and we get for the energy of the state $\psi = a^+_{mJ}|0\rangle$
$$
E_{mJ} = J +\sqrt{1+\l m^2} \ .
$$
This formula demonstrates explicitly that the energy of a long string with non-vanishing winding number
scales as $\sqrt{\l}$ contrary to the usual $\l^{1/4}$ scaling of a short string.

Eq.({\ref{ener3}) cannot be solved exactly for other states but it can be
readily used to compute $1/\sqrt{\l}$ corrections to the energy
of a long string. The form of the correction depends on
what scaling we assume for $J$, and mode numbers $n_i$. To illustrate the $J$ and mode number dependence
we present below a formula for the energy of a state obtained by acting by $M$ creation operators on the vacuum:
$$|m_1 J, \ldots ,m_M J\rangle = a^+_{m_1J}\cdots a^+_{m_MJ}|0\rangle\ ,$$
where $m_1 + m_2 +\cdots +m_M = m$. We assume that $J$ is kept fixed in the large $\l$ expansion.
Note also that $m_i$ can be positive and/or negative, and are not required to be integer. Then, by using
({\ref{ener3}) we find the energy of the state
\bea\nonumber
E_{m_1 J, \ldots ,m_M J} = m\sqrt{\l} &+& J \sum_{i=1}^M {|m_i|\ov m} +
{J^2\ov 2m\sqrt{\l}}\left( 1 - \left(\sum_{i=1}^M {|m_i|\ov m}\right)^2\right)\\\nonumber &+&
{1\ov 4\sqrt{\l}}\left(1 + \sum_{i=1}^M {|m_i|\ov m}\right)\sum_{j=1}^M {1\ov |m_j|} +\cdots
\eea
We see that for all these states the large $\l$ behavior is the same: $m\sqrt{\l}$. It is interesting that if some of $m_i$ are negative the constant term in the expansion is not equal to $J$. If all $m_i$ are positive
the formula simplifies and takes the form
\bea\nonumber
E_{m_1 J, \ldots ,m_M J} = m\sqrt{\l} + J
+{1\ov 2\sqrt{\l}}\sum_{j=1}^M {1\ov m_j} +\cdots \ ,\qquad m_i>0\ .
\eea

\section{ Relation to quantum string Bethe ansatz }
In this section we discuss the relation of the string
spectrum (\ref{ener3}) we obtained in the previous
section with the spectrum that can be derived by using
the quantum string Bethe ansatz for the $\su(1|1)$ sector.

We start by showing that eqs.(\ref{space-time}) and (\ref{ener3})
can be derived from the following set of Bethe ansatz type equations
\bea
\la{qsba}
{\rm exp}\left(ip_k L
+ {i\ov 2}\sum_{j=1}^M \left(p_k \left(e_j-1\right)-
\left(e_k-1\right) p_j\right)\right) =1\ .
\eea
Here $p_k$ are to be interpreted as the momenta of excitations of
a spin chain of length $L=J+M/2$ with $M$ excitations, and
\bea
\la{ek}
e_k = \sqrt{1+{\l p_k^2\ov 4\pi^2}}
\eea
is the energy of an elementary excitation, and the spectrum is determined
by the equation
\bea
\la{ene}
E - J = \sum_{k=1}^M  e_k
\eea
We have written eq.(\ref{ene}) in such a form to make obvious its similarity with
eq.({\ref{ener3}).

The sum over $j$ in eq.(\ref{qsba}) can be easily taken by using
(\ref{ene}), and we get \bea \la{qsba2} {\rm
exp}\left(ip_k\left(J+{1\ov 2}M\right) + {i\ov 2}\left(p_k
\left(E-J-M\right)-2\pi m\left(e_k-1\right) \right)\right) =1\,
,\eea where
$$
m={1\ov 2\pi}\sum_{j=1}^M p_j
$$
is the winding number which we suppose to be an integer.

Collecting the terms with $p_k$ together, we get
\bea
\la{qsba3}
{\rm exp}\left({i\ov 2}p_k\left(E+J\right) -
i\pi m e_k+i\pi m \right) =1\ .
\eea
Finally, taking the logarithm of both sides of the equation and
recalling that $P_+=E+J$, we obtain
\bea
\la{qsba4}
{1\ov 2}p_kP_+ -
\pi m e_k = 2\pi n_k\ ,
\eea
where the mode numbers $n_k$ are integer if the winding number $m$ is even, and
half-integer if $m$ is odd. Note that it is in complete agreement
with our consideration
in the previous section. The mode numbers $n_k$
cannot be arbitrary, they must satisfy a consistency
condition which should be equivalent to the level-matching condition.
To find the condition we take the sum over $k$ of both sides of
(\ref{qsba4}) and get
\bea
\la{con}
\sum_{k=1}^M  n_k = mJ\ .
\eea
It is exactly the same relation we obtained in
the previous section from the level-matching
condition.

To derive eqs.(\ref{space-time}) and (\ref{ener3}) let us first consider the
simplest case of the vanishing winding number. Then we get from
(\ref{qsba4}) \bea \la{qsba5} p_k = {4\pi n_k\ov E+J}\ , \eea and
substituting the formula into (\ref{ek}) and (\ref{ene}), we
immediately obtain eq.(\ref{space-time}).

The consideration can be easily generalized to the case of the
non-vanishing winding number. Expressing now $p_k$ as a function
of $e_k$, and solving eq.(\ref{qsba4}) for $e_k$, we get that the
energy of an elementary excitation $e_k$ with the mode number
$n_k$ coincides with the frequency $\omega_{n_k}^+$ (\ref{oma}),
and therefore eq.(\ref{ene}) just takes the form of
eq.(\ref{ener3}).

Thus, we
have shown that the space-time energy spectrum of strings
in the $\su(1|1)$ sector and in the uniform light-cone gauge follows
from the Bethe type equations (\ref{qsba}-\ref{ene}).
It is not difficult to see that the equations (\ref{qsba}) in fact coincide with
the Bethe ansatz equations for strings in the $\su(1|1)$ sector derived
in \cite{Staudacher:2004tk} by analyzing the near BMN spectrum.\footnote{Let
 us note that
in \cite{Staudacher:2004tk} the winding number $m$ was set to
zero. Our consideration here is a generalization of
\cite{Staudacher:2004tk} to the general $m$ case.} The only
difference is that in \cite{Staudacher:2004tk} the energy of an
elementary excitation was supposed to be \bea \la{ekqsba}
e_k=\sqrt{1+{\lambda\ov \pi^2} \sin^2{p_k\ov 2}}\ . \eea Comparing
this formula with eq.(\ref{ek}), we see that the set of
eqs.(\ref{qsba}-\ref{ene}) is an approximation of the quantum
string Bethe ansatz \cite{Staudacher:2004tk} valid in the regime
of small momenta $p_k$. According to (\ref{qsba5}), this is also a
high-energy regime where $n_i/(E+J)\ll1$. It is not difficult to
see that in the large $J$ limit the approximation is valid up to
the $1/J$ order (see appendix A for details) and in the strong
coupling limit only up to the leading $\l^{1/4}$ order. At higher
orders in $1/J$ and $1/\l^{1/4}$ the $\sin^2\sfrac{p}{2}$ begins
to give additional contributions to the space-time energy. That
means that computing $1/J^2$ and $1/\l^{1/4}$ corrections should
provide nontrivial tests of the dispersion relation
(\ref{ekqsba}). Let us stress out that in \cite{Arutyunov:2004vx} the choice of
the dispersion relation (\ref{ekqsba})
was  motivated by the asymptotic Bethe ansatz in gauge theory \cite{Beisert:2004hm}.
A priori there is no reason why it should not be corrected at large $\l$.
Recent tests of the quantum string Bethe ansatz
for the $\su(2)$ and $\sls(2)$ sectors performed in
\cite{SZZ,MTT} indicate, however, that $1/J^2$ corrections are compatible
with the dispersion relation. It would be very interesting to
compute the leading $1/\l^{1/4}$ corrections.

\section{Conclusion}
In this paper we have introduced the uniform light-cone gauge for
superstrings in $\AdS$, and applied it to analyze the classically-consistent
reduction of the Green-Schwarz action to the $\su(1|1)$ sector.
\medskip

It appears that in this gauge the reduced model is described by a free
action of two complex fermions, and therefore the spectrum of the
model can be easily found. We have explained how the spectrum is related
to the space-time energy spectrum of strings that by the AdS/CFT correspondence
coincides with the spectrum of scaling dimensions of ${\cal N}=4$ SYM.
\medskip

The space-time energy spectrum appeared to reproduce correctly the leading
$1/J$ correction in the large $J$ limit, and the leading $\l^{1/4}$ behavior in
the strong coupling limit. We have shown that
the space-time energy equation (\ref{ener3})
can be reproduced from the low-momentum approximation to the quantum string Bethe
ansatz for the $\su(1|1)$ sector.

\medskip
We have noted however that the naive $1/J^2$ and $1/\sqrt{\l}$
corrections found by using the free fermion action differ from the
predictions of the quantum string Bethe ansatz, and computing them
by using the whole string sigma-model would provide nontrivial
tests of the ansatz.

\medskip
Calculating the $1/J^2$ corrections and even the leading
$1/\sqrt{\l}$ corrections would require using the second-order
perturbation theory. The uniform light-cone gauge and $\su(1|1)$
sector seem to be the most suitable ones for such a computation
because in this gauge the sector is described by a free theory,
and therefore only the fields that were set to zero in the
reduction to the  $\su(1|1)$ sector would contribute to the
corrections.

\medskip
Our results also show explicitly that quantizing a
classically-closed sector of superstrings in $\AdS$ cannot lead to
results correct for finite $J$ and $\l$. Moreover, quantizing such
a sector in different gauges might lead to contradictory results.
The reason for that is that the quantum gauge equivalence requires
the conformal invariance of the string theory that is broken when
we reduce the theory to a classically-closed sector. It would be
interesting to study the gauge dependence of the $\su(1|1)$ sector
spectrum by quantizing the integrable model of a massive Dirac
fermion \cite{AAF} that describes strings in the $\su(1|1)$ sector
in the uniform gauge $p_\p=J$.
\medskip

We conclude therefore that the results derived within a closed
sector must be taken with great care since the correct
quantization of superstrings in $\AdS$ would require taking into
account all bosonic and fermionic fields of the superstring.

\newpage

\section*{Acknowledgments}
We are grateful to  F.~Alday, N.~Dorey, J.~Plefka, M.~Staudacher,
A.~Tseytlin and M.~Zamaklar for many valuable discussions. G.~A.
would like to thank Emery Sokatchev and Laboratoire
d'Annecy-Le-Vieux de Physique Th\'eorique, where this work was
completed, for the kind hospitality. The work of G.~A. was
supported in part by the European Commission RTN programme
HPRN-CT-2000-00131 and by RFBI grant N02-01-00695. The work of
S.~F.~was supported in part by the EU-RTN network {\it
Constituents, Fundamental Forces and Symmetries of the Universe}
(MRTN-CT-2004-005104).

\appendix

\section{Quantum String Bethe Ansatz in the $\su(1|1)$ sector}
Here we outline the derivation of the $1/J^2$ correction to the
energy of a plane-wave state by using the conjectured quantum
string Bethe ansatz
in the $\su(1|1)$ sector \cite{Staudacher:2004tk,Beisert:2005fw}.
We will also analyze the strong coupling expansion
$\lambda\to\infty$.

\subsection{The $1/J$ expansion}

To formulate the Bethe equations one introduces the function
$$
x(u)=\sfrac{1}{2}u+\sfrac{1}{2}u\sqrt{1-\frac{2g^2}{u^2}}\, ,
~~~~~~~~~g^2=\frac{\lambda' J^2}{8\pi^2}\,
$$
and the notation $x^{\pm}=x(u\pm \sfrac{i}{2})$. The variable $u$
is related to the momentum $p$ of an elementary excitation through
the formula \bea ip=\log \frac{x^+(u)}{x^-(u)} \, .\label{mom}\eea
The (logarithm of) Bethe equations are the set of $M$ equations
for the momenta $p_k$, $k=1,\ldots, M$. In the $\su(1|1)$ sector
they read \cite{Beisert:2005fw} \bea &&iLp_k=2\pi i
n_k+\sum_{j\neq k}^M \log\big[1-\frac{g^2}{2x_k^-x_j^+}\big]
-\log\big[1-\frac{g^2}{2x_k^+x_j^-}\big]\\
\nonumber
&&+iu_{kj}\Big[\log\big[1-\frac{g^2}{2x_k^-x_j^+}\big]+\log\big[1-\frac{g^2}{2x_k^+x_j^-}\big]
-\log\big[1-\frac{g^2}{2x_k^+x_j^+}\big]-\log\big[1-\frac{g^2}{2x_k^-x_j^-}\big]\Big].
\eea Here $n_k$ are the excitation numbers, $u_{kj}\equiv u_k-u_j$
and $L$ is the {\it length} which in our present case is
$L=J+\sfrac{1}{2}M$. As soon as momenta $p_k$ are found the energy
can be computed by using the formula
\bea \label{ener} E^{\rm
BA}=ig^2\sum_{k=1}^M
\left(\frac{1}{x_k^+}-\frac{1}{x_k^-}\right)=\sum_{k=1}^M
\Big[-1+\sqrt{1+\frac{\lambda}{\pi^2}\sin^2\sfrac{p_k}{2}
}\Big]\, . \eea

\medskip
Assuming the following expansion for momentum $p_k$ in the large
$J$ limit
 \bea p_k=\frac{2\pi
n_k}{J}+\frac{p_k^{(2)}}{J^2}+\frac{p_k^{(3)}}{J^3}+\ldots \eea we
determine the leading behavior of $u_k\equiv u(p_k)$ by using the
formula (\ref{mom}). We find $$ u_k=\frac{J\omega_k}{2\pi
n_k}-\frac{p_k^{(2)}}{4\pi^2\omega_kn_k^2}+\frac{(3\omega_k^2-1)
(-4\pi^4n_k^2\omega_k^2+3(p_k^{(2)})^2)-12\pi
n_k \omega_k^2 p^{(3)}_k}{48\pi^3n_k^3\omega_k^3} +\ldots $$ The
Bethe equations generate then the perturbative solution for $p_k$
\bea\label{m1} \frac{p_k^{(2)}}{\pi}=- M n_k + \sum_{j\neq k}^M
n_k(1-\omega_j)-n_j(1-\omega_k) \eea and \bea \label{m2}
p_k^{(3)}=-\frac{1}{2}Mp_k^{(2)}+\frac{1}{2} \sum_{j\neq k}^M
p_j^{(2)}\left(\omega_k-1-\frac{\lambda'n_jn_k}{\omega_j}\right)
-p_k^{(2)}\left(\omega_j-1-\frac{\lambda'n_jn_k}{\omega_k}\right)\,
. \eea Now by using eq.(\ref{ener}) we obtain the first few
leading terms in the large $J$ expansion of the energy
$$
E^{\rm BA}=\sum_{k=1}^M (\omega_k-1)+\frac{E_1^{\rm
BA}}{J}+\frac{E_2^{\rm BA}}{J^2}+\ldots ,
$$
where \bea \label{E1ba} E_1^{\rm BA}=\lambda'\sum_{k=1}^M
\frac{n_k}{\omega_k}\frac{p_k^{(2)}}{2\pi} \eea and \bea
\label{BAp} E_2^{\rm
BA}=\frac{\lambda'}{24}\sum_{k=1}^M\frac{1}{\omega_k^3}\left[3
\left(\frac{p_k^{(2)}}{\pi}\right)^2+
12n_k\omega_k^2 \left(\frac{p_k^{(3)}}{\pi}\right)
-4\pi^2n_k^4\omega_k^2\right]\, . \eea Thus, we have computed the
$1/J^2$ correction $E_2$ to the energy of the plane-wave
$M$-impurity state. On the other hand, the theory of the free
Dirac fermion leads to the corrections (\ref{ppw}) and
(\ref{ppw2}) which we repeat here for convenience \bea
E_1&=&-\frac{\lambda'}{2}\sum_{i,j=1}^M \omega_i \frac{n_j^2}{\omega_j}\, ,\\
\label{ppw21} E_2 &=&{\l'\ov 8} \sum_{i,j,k=1}^M\omega_i
\omega_j{n_k^2(3+2\l' n_k^2)\ov \omega_k^3} + {\l'^2\ov 4}
\sum_{i,j,k=1}^M\omega_k{n_i^2n_j^2 \ov \omega_i\omega_j} \, .\eea
Upon substituting in eqs.(\ref{E1ba}),(\ref{BAp}) the momenta
(\ref{m1}) and (\ref{m2}) one can show that the first correction
to the energy is the one and the same, while eq.(\ref{BAp})
coincides with eq.(\ref{ppw21}) except for the term which
explicitly depends on $\pi^2$: \bea \nonumber E_1^{\rm BA}&=&E_1\, , \\
E_2^{\rm BA}&=&
E_2-\pi^2\frac{\lambda'}{6}\sum_{k=1}^M\frac{n_k^4}{\omega_k}\, .
\eea Thus, the formula for $E_2^{\rm BA}$ {\it disagrees} with
eq.(\ref{ppw2}) which
 provides the $1/J^2$ correction from
the theory of the free Dirac fermion. In fact, the additional term
proportional to $\pi^2$ in the expression for the Bethe ansatz
energy $E_2^{\rm BA}$ occurs due to the term $p^4$ of
$\sin^2\sfrac{p}{2}$ in the large $J$ expansion of the elementary
excitation charge $e_k$ in eq.(\ref{ekqsba}).

\subsection{The strong coupling expansion}

Now we analyze the corrections to the strong coupling limit
$\lambda\to \infty$, $L\sim 1$. Our discussion is very close to
that of \cite{Arutyunov:2004vx}. To investigate the strong
coupling expansion  it is convenient to express the $x_{\pm}$ as
functions of the momentum $p$.
\bea x_{\pm}(p)&=&\frac{e^{\pm
i\frac{p}{2}}}{4\sin\frac{p}{2}}\Big(1+\sqrt{1+\frac{\lambda}{\pi^2}\sin^2\frac{p}{2}}~\Big)\,
.
\eea
The function $u(p)$ is then
\bea
u(p)=\sfrac{1}{2}\cot(\sfrac{p}{2})\sqrt{1+\frac{\lambda}{\pi^2}\sin^2\frac{p}{2}}\,
. \eea We assume that in the strong coupling regime the momentum
$p$ admits the following expansion \bea p=\frac{p^{(1)}
}{\sqrt[4]{\lambda} }+\frac{p^{(2)}}{\sqrt{\lambda}}+\ldots ,\eea
where the coefficients $p^{(i)}$ should be determined from the
Bethe ansatz equations. The roots $p_k$ (generically complex) obey
the conservation law $\sum_{k=1}^Mp_k=0$. We find it convenient to
group the leading momenta $p_k^{(1)}$ into two sets: $p_k^+$ with
${\rm Re} p_k^{(1)}>0$, $k=1,\ldots, m$ and $p_k^-$ with ${\rm Re}
p_k^{(1)}<0$, $k=m+1,\ldots, M$.

\medskip

Expanding the Bethe equations in the limit $\lambda\to \infty$ we
obtain at the first three leading orders the following equations
\bea \label{blead} &&~~\lambda^{0}:~~~ 2\pi
n_k-\frac{1}{2\pi}\sum_{j=m+1}^M
p^{+}_kp^{-}_j+\sum_{j=1}^m\chi_{kj}^{(1)}=0\, , \\
\label{sub} &&\frac{1}{\sqrt[4]{\lambda} } :~~~
Lp_k^+=\frac{1}{2\pi}\sum_{j=m+1}^M \Big[\pi(p_k^+
-p_j^-)+p_k^+p_j^{(2)}+p_j^-p_k^{(2)}\Big]
+\sum_{j=1}^m\chi_{kj}^{(2)}\, ,\
\\
\nonumber &&\frac{1}{\sqrt{\lambda} } :~~~
Lp_k^{(2)}=\sum_{j=m+1}^M
 \Big[
\frac{\pi}{2}\Big(\frac{p_j^-}{p_k^+}+\frac{p_k^+}{p_j^-}\Big)
-\frac{1}{48\pi}(p_j^{-3}p_k^++p_j^-p_k^{+3})+\frac{1}{2}(p_k^{(2)}-p_j^{(2)})+\\
\label{subsub}
 &&~~~~~~~~~~~~~~~~~+\frac{1}{2\pi}p_j^{(2)}p_k^{(2)}
+\frac{1}{2\pi}(p_k^+p_j^{(3)}+p_j^-p_k^{(3)}) \Big]+\sum_{j=1}^m
\chi^{(3)}_{kj}\, . \eea Here we presented equations for
$k=1,\ldots ,m$ only as they are enough for our further
discussion. The functions $\chi_{kj}^{(1)}$ are rather
complicated, e.g., \bea \chi^{(1)}_{kj}&=&
\frac{1}{i}\log\frac{\frac{1}{p_j^+}+\frac{1}{p_k^+}+\frac{i}{4\pi}(p_j^+-p_k^+)
}{\frac{1}{p_j^+}+\frac{1}{p_k^+}-\frac{i}{4\pi}(p_j^+-p_k^+) } +\\
\nonumber &+&2\pi
\Big[\frac{1}{p_k^{+2}}-\frac{1}{p_j^{+2}}-\frac{1}{16\pi^2}(p_k^{+2}-p_j^{+2})\Big]
\log\frac{\Big(\frac{1}{p_j^+}+\frac{1}{p_k^+}\Big)^2
+\frac{1}{16\pi^2}(p_j^+-p_k^+)^2
}{\Big(\frac{1}{p_j^+}+\frac{1}{p_k^+}\Big)^2+\frac{1}{16\pi^2}(p_j^++p_k^+)^2
} \, . \eea Fortunately, the only property of $\chi_{kj}^{(i)}$ we
need, is that they are antisymmetric w.r.t. the change
$i\leftrightarrow j $: $\chi_{kj}^{(i)}=-\chi_{jk}^{(i)}$.

\medskip

Summing up the Bethe equations (\ref{blead}) over $k$ and using
the antisymmetry property of $\chi_{kj}^{(1)}$ we first find \bea
\sum_{k=1}^mp_k^+=-\sum_{k=m+1}^Mp_k^-=2\pi \sqrt{n}\, ,
~~~~~~~n\equiv \sum_{k=1}^m n_k\, . \label{fp} \eea Second,
summing up eqs.(\ref{sub}) and using the momentum conservation we
obtain \bea \label{l2}
L-\frac{M}{2}=-\frac{1}{2\pi}\left(\sum_{k=1}^mp_k^{(2)}-\sum_{j=m+1}^Mp_j^{(2)}\right)\,
. \eea Third, summing up eqs.(\ref{subsub}) and using
eqs.(\ref{fp}),(\ref{l2}) we get the relation \bea \nonumber
\frac{\Big(L-\frac{M}{2}\Big)^2}{4
\sqrt{n}}=\sum_{k=1}^m\frac{24\pi^2-p_k^{+4}+24p_k^+p_k^{(3)}}{48\pi
p_k^+}
-\sum_{j=m+1}^M\frac{24\pi^2-p_j^{-4}+24p_j^-p_j^{(3)}}{48\pi
p_j^-}\, . \eea

\medskip

Expanding the energy in the large $\lambda$ limit we get \bea
\nonumber E^{\rm
BA}&=&-M+\frac{\sqrt[4]{\lambda}}{2\pi}\left(\sum_{k=1}^mp_k^+-\sum_{k=m+1}^Mp_k^-\right)
+\frac{1}{2\pi}\left(\sum_{k=1}^mp_k^{(2)}-\sum_{j=m+1}^Mp_j^{(2)}\right)+\\
&+&\frac{1}{\sqrt[4]{\lambda}}\left(\sum_{k=1}^m\frac{48\pi^2-p_k^{+4}+24p_k^+p_k^{(3)}}{48\pi
p_k^+}
-\sum_{j=m+1}^M\frac{48\pi^2-p_j^{-4}+24p_j^-p_j^{(3)}}{48\pi
p_j^-}\right)+\,\ldots \nonumber \eea Substituting here our
findings we obtain
$$
E^{\rm BA}=2\left(n^2\lambda\right)^{\frac{1}{4}}
-\Big(L+\frac{1}{2}M\Big)+\frac{1}{\sqrt[4]{\lambda}}\left[\frac{\Big(L-\frac{M}{2}\Big)^2}{4
\sqrt{n}}+\frac{\pi}{2}\Big(\sum_{k=1}^m\frac{1}{p_k^+}-\sum_{j=m+1}^M\frac{1}{p_j^-}\Big)\right]
+\ldots \,
$$
It is rather remarkable that the subleading term in the strong
coupling expansion of the Bethe ansatz energy appears to coincide
with the canonical dimension of the gauge theory operator taken
with the negative sign. Thus, at strong coupling the total
conformal dimension, $\Delta=J+M+E^{\rm BA}$, of the dual gauge
theory operator has an expansion \bea \label{strong}
\Delta=2\left(n^2\lambda\right)^{\frac{1}{4}}
+\frac{1}{\sqrt[4]{\lambda}}\left[\frac{J^2}{4
\sqrt{n}}+\frac{\pi}{2}\Big(\sum_{k=1}^m\frac{1}{p_k^+}-\sum_{j=m+1}^M\frac{1}{p_j^-}\Big)\right]
+\ldots \,\, . \eea Thus, as in \cite{Arutyunov:2004vx}, we found
that the gauge theory operators are dual to string modes with
masses $m^2=4n\sqrt{\lambda}$, where the level $n$ is determined
by the mode numbers of the roots with a positive real part
$n=\sum_{k=1}^m n_k$. We also see that the constant term in the
large $\lambda$ expansion {\it cancels out}. We were not able to
write this formula for general $M$ in terms of the excitation
numbers $n_k$ because  the individual momenta $p_k$, which are
solutions of the complicated equation (\ref{blead}), are not
explicitly known.

\medskip

The formula (\ref{strong}) can be confronted with
eq.(\ref{strong1}) describing the strong coupling asymptotics of our
reduced model. For the reader convenience we repeat this equation
here \bea \label{strongrepeat}
E=2\left(n^2\lambda\right)^{\frac{1}{4}}
+\frac{1}{\sqrt[4]{\lambda}}\left[\frac{J^2}{4
\sqrt{n}}+\frac{\sqrt{n}}{4}\sum_{i}\frac{1}{|n_i|}\right] +\ldots
\,\, .  \eea The leading terms in eqs.(\ref{strong}) and
(\ref{strongrepeat}) coincide. In both equations the constant
(subleading) piece is absent. However, the terms of order
$1/\sqrt[4]{\lambda}$ are different. They coincide only for the
special case of two impurities, $M=2$.

\medskip

The results about the subleading behavior of conformal dimensions
should be taken with caution. Indeed, in general the quantum
string Bethe ansatz \cite{Arutyunov:2004vx} involves an infinite
number interpolating functions $c_r(\lambda)$ with the property
$c_r(\lambda)\to 1$ as $\lambda\to\infty$. The results by
\cite{BT} suggest that beyond the semiclassical limit these
functions become non-trivial. In our computation above we assumed
that all $c_r=1$. Taking into account the actual functions $c_r$
(yet to be determined) might change the conclusion about the
strong coupling expansion beyond the leading order.

\section{Lax representation}

Integrability of the classical superstring theory on $\AdS$ was
demonstrated in \cite{Bena:2003wd} by means of constructing the
Lax (zero-curvature) representation for the superstring equations
of motion. In \cite{AAF} we have shown that this connection admits
a consistent reduction to the fields describing excitations from
the $\su(1|1)$ sector. Thus, the non-trivial interacting Dirac
Hamiltonian \cite{AAF} which governs the dynamics in this sector
is integrable, but its integrable properties are not transparent
rather they are hidden in the highly non-trivial Lax pair. This
pair can be formulated in terms of two $4\times 4$ matrices,
$\L_{\sigma}$ and $\L_{\tau}$, depending on a spectral parameter
$z$ and satisfying the condition of zero curvature
 \bea \label{zc}
\pa_{\sigma}\mathscr{L}_{\tau}-\pa_{\tau}\mathscr{L}_{\sigma}-
[\mathscr{L}_{\sigma},\mathscr{L}_{\tau}]=0 \eea as a consequence
of the dynamical equations.

\medskip

In opposite, in our present approach based on the uniform
light-cone gauge integrability of the reduced model is manifest as
it is the theory of a  free 2-dim massive  Dirac fermion. In fact
the proper choice of the gauge resulted into linearization of the
dynamical equations! Non-triviality of the original theory is now
hidden in the new dispersion formula relating the energy $E$ to
the charges $M$ and $J$.

\medskip

In spite of the manifest integrability of the model it is still
interesting to know what is the reduction of the general Lax
connection to the $\su(1|1)$ sector in the uniform light-cone
gauge. For the sake of clarity we restrict our further discussion
to the case of zero winding number\footnote{Of course, for $m\neq
0$ the reduced Lax connection also exists.}, $m=0$, and fix
$\kappa=\sfrac{\sqrt{\lambda}}{2}$. We also introduce the concise
notation for the original fermionic variables $\vartheta$ of
\cite{AAF} \bea \psi_1=\vartheta_3\, , ~~~~\psi_2=\vartheta^8\, ,
~~~~ \stackrel{*}{\psi_1}=\vartheta^3\, ,
~~~~\stackrel{*}{\psi_2}=\vartheta_8\, \eea as well as two even
quantities \bea
\varsigma&=&\psi_1\stackrel{*}{\psi'_1}+\stackrel{*}{\psi_1}\psi'_1-
\stackrel{*}{\psi_2}\psi'_2-\psi_2\stackrel{*}{\psi'_2}\, ,\\
\varrho&=&(\psi_1\stackrel{*}{\psi_2}-\stackrel{*}{\psi_1}\psi_2)'\,
. \eea

One can further show that the minimal Lax connection for the
$\su(1|1)$ sector in the uniform light-cone gauge is given in
terms of $2\times 2$ matrices of the form \bea \L_{\s}&=&\left(
\begin{array}{cc} -\frac{i}{\sqrt{\tilde{\lambda}}}\frac{z}{1-z^2}+
\frac{1}{4}\varsigma ~~&~~
\frac{-\stackrel{*}{\psi'_1}-iz\stackrel{*}{\psi'_2}}{\sqrt{1-z^2}} \\
~~&~~ \\
\frac{-\psi'_1+iz\psi'_2}{\sqrt{1-z^2}} ~~&~~
\frac{i}{\sqrt{\tilde{\lambda}}}\frac{z}{1-z^2}+\frac{1}{4}\varsigma
\end{array}\right)\, ,\\
\L_{\tau}&=&\left(  \begin{array}{cc}
-\frac{i}{2}\frac{1+z^2}{1-z^2}+\frac{i\sqrt{\tilde{\lambda}}}{4}\varrho
~~&~~
-\sqrt{\tilde{\lambda}}\frac{z\stackrel{*}{\psi'_1}+
i\stackrel{*}{\psi'_2}}{\sqrt{1-z^2}} \\
~~&~~ \\
-\footnotesize{\sqrt{\tilde{\lambda}}}\frac{z\psi'_1-i\psi'_2}{\sqrt{1-z^2}}
~~&~~
\frac{i}{2}\frac{1+z^2}{1-z^2}+\frac{i\sqrt{\tilde{\lambda}}}{4}\varrho
\end{array}  \right)\, . \eea
Here $z$ is the spectral parameter and we recall the definition
$\tilde{\lambda}=\frac{4\lambda}{P_+^2}$. One can easily check
that the Lax connection above has  zero curvature (\ref{zc}) by
virtue of the fermionic equations of motion followed from
eq.(\ref{gxL}) \bea \label{evf}
\dot{\psi}_1=i\psi_1-i\sqrt{\tilde{\lambda}}~\psi'_2\, , ~~~~~~~
\dot{\psi}_2=-i\psi_2+i\sqrt{\tilde{\lambda}}~\psi'_1\, . \eea
Thus, the Lax connection for the whole superstring sigma-model
\cite{Bena:2003wd} boils down under the reduction to the
$\su(1|1)$ sector in the uniform light-cone gauge to that of the
free massive Dirac fermion. In writing component $\L_{\tau}$ we
also used the evolution equations (\ref{evf}) to trade the
$\tau$-derivatives of fermions for their $\sigma$-derivatives.

\medskip

Some comments are in order. Upon choosing the minimal reduction to
$2\times 2$ matrices the corresponding Lax connection has the
non-vanishing supertrace (its usual trace is also non-zero). This
is not problematic since the only requirement that matters is the
fulfillment of the equations of motion which is indeed the case.
One can also see that the off-diagonal part of $\L_{\sigma}$
contains the $\sigma$-derivatives of the fermions rather than the
fermions themselves. This problem can be cured by means of a
certain gauge transformation which  removes  in $\L_{\sigma}$ the
off-diagonal derivatives in favor of the fields. Then the
off-diagonal elements of eq.(\ref{zc}) will directly generate the
fermionic equations (\ref{evf}) rather than their
$\sigma$-derivatives. This gauge transformation leads however to a
slightly more complicated form of the diagonal matrix elements in
$\L_s$ and $\L_{\tau}$ and, therefore, we have not attempted to
discuss it here.


\end{document}